# TWO NEW VORTEX LIQUIDS

Abstract:  It is suggested that the observations of nonlinear susceptibility and Nernst effect in cuprate superconductors above Tc, and those of non-classical rotational inertia in solid He, are two manifestations of a state of matter we call a "vortex liquid", distinct from a conventional liquid in that its properties are dominated by conserved supercurrents flowing around a thermally fluctuating tangle of vortices.

In the past two years we have been confronted with puzzling data involving  "non-classical" responses to vorticity in two very different physical systems.  One has garnered (justifiably) a great deal of  attention: the observations by Moses Chan and co-workers[1] of NCRI (Leggett's term[2] : Non-Classical Rotational Inertia) in samples of solid He-4 and solid $H_2$.

Attention for the second has been confined mostly to specialists in the field of high Tc superconductivity.  In these superconductors there is a phase for T>Tc, in which anomalous properties persist up to quite high temperatures—it is called the "pseudogap" phase. It has recently been shown[3] that in the lower portion of this phase the in-plane diamagnetic susceptibility is large and nonlinear up to a certain fairly definite "onset" field and temperature, lower than the crossover T* which bounds the pseudogap region.  That this behavior is due to vortex-like currents is demonstrated by the earlier observation[4] of a large and nonlinear Nernst effect, continuous with that observed in the high-field flux-flow region, and ascribable to viscous flow of vortices along a thermal gradient. Above Tc the susceptibility and Nernst signal track each other closely.

The moment of inertia can be thought of as a rotational susceptibility

$$I = \frac{\partial M}{\partial \Omega}$$

, where M is angular momentum and Ω angular velocity. Thus nonlinear susceptibility and NCRI are similar phenomena. The two experimental examples can be shown to be similar even in functional form if we note that M/B often varies as lnB over an intermediate range of B, while δI falls off linearly with ln Ω at larger Ω. (ref 1)

I believe that in both of these phases the dynamics are controlled by thermally excited, fluctuating, quantized vortex tangles: they are vortex liquids. The concept of vortex liquid is not a new one: it is implicit in the Kosterlitz-Thouless[5] theory of the superfluid phase transition in 2 dimensions, where the superfluid phase transition occurs via the thermal proliferation of vortices. The melting of the Abrikosov lattice of flux lines is known to be the transition mechanism at higher fields in high Tc superconductors[6]. The vortex liquid is a relatively familiar concept in this high field range, and there have been suggestions that it is a distinct phase[7], but prior to our recent work there has not been a very clear characterization of it or its dynamics. We are suggesting that this entire region in the field-temperature phase diagram is not just some consequence of either superconducting fluctuations in a basically normal fluid, or of a melted array of Abrikosov vortices. We may formally characterize it by its nonanalytic response to magnetic field, a response which vanishes outside of a definite region of temperature and field. Sudbo and Nguyen[8] have also suggested the existence of a distinct vortex liquid phase above the superfluid transition in their simulations, but have not discussed dynamics.

The transition temperature Tc for a superfluid is normally described in terms of the x-y model, where the order parameter is a phase $\phi$ and the effective Hamiltonian is proportional to $(\nabla\phi)^2$. This seems also to be the case for the high Tc superconductors, empirically[9] and now theoretically[10]; the energy gap persists for a considerable range above Tc, unlike conventional superconductors, and Tc is caused by phase disordering. There is good empirical reason to believe (ref 10) that quasiparticle currents are dominated by the supercurrents at least near Tc.

The x-y model, however, does not really describe the disordered superfluid, because the order parameter is not a simple classical vector but a quantum dynamical variable. There are supercurrents whose velocity is given by

$$v = \frac{\hbar}{m}\nabla\phi \text{ and current by } J = \rho_s v, \qquad [1]$$

The crucial concept here is that there is a finite superfluid density $\rho_s$ which has an equilibrium value at any temperature below the onset and at any field below $H_{c2}$, hence there is a term in the free energy

$$F = const \times (\rho_s - \rho_{s0})^2 \qquad [2]$$

This ensures that the superfluid is approximately incompressible on the large time and space scale of phase fluctuations, so that

$$\nabla \cdot J_s = 0 = \nabla^2 \phi, \qquad [3]$$

coarse-grained over such scales. (Our $\rho_s$ is not the conventional macroscopic quantity giving the penetration depth, which is its average over phase fluctuations, but is defined microscopically by equation [1].) This means that the state of the vortex fluid at any given instant can be completely characterized by an array of singular lines (vortices) around which the phase rotates by $2\pi$. In the homogeneous fluid or in the absence of strong pinning forces these vortex lines will move with the local velocity of the fluid due to all the other lines. I have been able[11] to give a reasonably good quantitative account of the Nernst observations on the pseudogap phase of high Tc superconductors, as well as of the resistivity in this phase, using this vortex fluid picture. The essential point of these transport phenomena is that dissipation is caused by the thermal fluctuations of the supercurrents with a mean correlation time $\tau \cong h/kT$. The observations of "Fermi Arcs" by Campuzano[12] in the pseudogap phase also receives a simple explanation in terms of the time-fluctuating phase of the energy gap.

The key property of the vortex fluid phase is that as long as the phase $\phi$ is definable and satisfies the constraint [2], the extra vortices caused by overall rotation of the sample, or equivalently by an external magnetic field in the superconductor case, cannot be screened away by polarization of the background thermal fluid of vortices. For each quantum unit of flux or circulation, there will be exactly one extra quantized vortex running entirely through the sample with its extra rotation by $2\pi$ along any circumscribing path, and therefore it will have its extra current implied by [1],

$$J_\theta = \hbar/mr \quad [4]$$

Hence there will be a non-classical diamagnetism or rotational inertia. I believe that the condition of finite, conserved superfluid density is likely to be equivalent to that in reference 6, that the line tension of vortex lines be finite.

The evidence for vortex fluid behavior in solid He is even less accepted and the following description of the situation is somewhat conjectural. D Huse[13] first pointed out that the observations of Chan et al in reference [1] appear not to be what is expected of a true superfluid but rather represent the dispersion at a dissipation peak above the superfluid Tc, in analogy with the observations of Kosterlitz-Thouless behavior in He films.[14] We proposed, and supported with experimental evidence, the idea that solid He is an incommensurate three-dimensional density wave rather than a true solid, with the actual helium capable of flowing as a fluid through this density wave. In the presumed ground state, this flow is that of a superfluid but above Tc (which we estimate is <.02 K) it is a vortex fluid. This flow has a superfluid density

$$\rho_s = n\hbar^2/2M, \text{ with } M \approx 10^2 M_{He} \quad [5]$$

and velocity

$$v_s = \hbar\nabla\phi/M \quad [6]$$

The dynamical variable which characterizes the flow of the helium relative to the lattice is the phase ϕ(r), which in turn is entirely described by a network of vortex lines plus uniform motion. There are only one set of Goldstone bosons, the phonons of the lattice, and actual compressional modes of the relative motion with respect to the lattice are at high frequency. When the lattice is in uniform motion without the fluid it carries only (1-$M_{He}$/M) of the mass. For the entire solid to rotate rigidly it must contain a number of vortices equal to Ω$M_{He}$/h per unit area.

Our suggestion is that the torsional vibration frequency 1000 Hz matches the rate at which vortices can move into and out of the sample precisely at the temperature of maximum dissipation. ( See fig 1). At lower temperatures the rate of vortex motion is too slow and the moment of inertia is only that of the lattice, hence is reduced by the ratio 1-$M_{He}$/M. At higher temperatures vortices flow in easily and the moment of inertia is normal.

Estimating the rate of motion of vortices in the pure solid using the methods of ref 11 seems to result in too rapid motion. He3 impurities should act as quite efficient pinning centers, and this is the most likely mechanism for the observations by Chan et al that the effects are very He-3 dependent, and disappear when He3 is completely absent. It may be that the vortices would have to carry the He-3 along with them, which would slow their motion severely. Structural defects, if present, would be less effective in pinning vortices, although the fact that annealing affects the results suggests some role in slowing vortex motion.

The very close analogy between the Nernst effect and the Chan effect is striking. In both cases a current of vortices is driven, in the one case by the thermal gradient and in the other by the alternating torsional velocity. What is actually measured is the viscous resistance to this flow, and the size of the response is proportional to the logarithmic energy of the vortices. The number of driven vortices is enormously different, of course: a small integer in Chan's case, of order $10^9$ in the Nernst effect.

Structural defects are often suggested as an "explanation" of the Chan experiments. The coincidence that the magnitude as a function of velocity corresponds exactly to the number of vorticity quanta argues against this; also, it is hard for me to understand why structural defects should respond specifically to rotation, mimicking the effects of vortices, and not, for instance, to steady pressures such as have been applied in various unsuccessful experiments. As with a Josephson junction, D C superflow will only occur at zero pressure difference, where there is no vortex flow.

Clearly the crucial experiment for our hypothesis is to change the torsional vibration frequency, holding all other variables constant. This has not been done. It would seem to be urgent to do so, since no other hypothesis yet proposed is consistent with any appreciable fraction of the data.

In conclusion, I am proposing that the extensive observations of Ong and Wang on the pseudogap phase of cuprate superconductors constitute the discovery of the vortex fluid phase conjectured by

Feigel'man and by Nguyen and Sudbo; and I conjecture that the observation of NCRI in quantum solids by Chan constitutes a rediscovery of this phase in a uniquely interesting system.

ACKNOWLEDGEMENTS

I must acknowledge valuable discussions with my collaborators D Huse and W F Brinkman, and also with S Sondhi and V Oganesyan (the latter of whose ideas, together with those of S Ragu, suggested the absence of screening in the vortex liquid). None of the above need take any responsibility for my conclusions. Experimental information from M Chan, Y Wang, L Li, and many discussions of theory and experiment with N P Ong, have been vital.

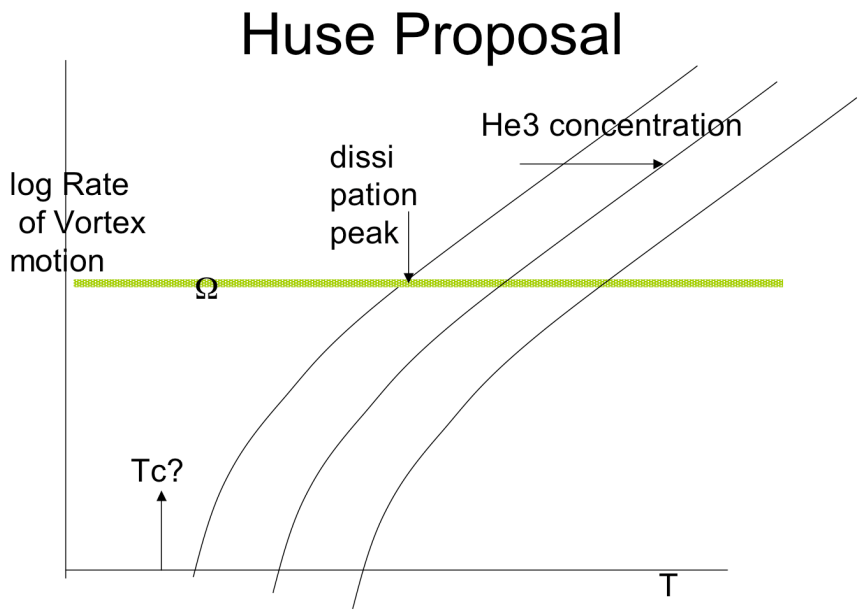

Figure 1